\documentclass[onecolumn]{aa}
\usepackage{graphicx}
\usepackage{natbib}
\usepackage{deluxetable}
\usepackage{txfonts}
\begin{document}
   \title{Multiplicity of very low-mass objects in the Upper Scorpius OB association: a possible wide binary population\thanks{Based on observations made with ESO Telescopes at the Paranal Observatory and with the William Herschel Telescope of the Isaac Newton Group of Telescopes in La Palma}}

   \author{H. Bouy\inst{1}
     \and E.~L. Mart\'\i n\inst{1,8}
     \and W. Brandner\inst{2}
     \and M. R. Zapatero-Osorio\inst{3}
     \and V.~J.~S.~B\'ejar\inst{4}
     \and M. Schirmer\inst{5}
     \and N. Hu\'elamo\inst{6}
     \and A.~M. Ghez\inst{7}
}

   \offprints{H. Bouy}

   \institute{Instituto de Astrof\'\i sica de Canarias, C/ V\'\i a L\'actea, s/n, E-38200 - La Laguna, Tenerife, Spain\\
              \email{bouy@iac.es, ege@iac.es}
	 \and
	     Max-Planck Institut f\"ur Astronomie, K\"onigstuhl 17, D-69117 Heidelberg, Germany\\
	     \email{brandner@mpia.de}
	 \and
	     LAEFF-INTA, PO50727, E-28080 Madrid, Spain\\
	     \email{mosorio@laeff.esa.es}
	 \and
	     Instituto de Astrof\'\i sica de Canarias, GTC project, C/ V\'\i a L\'actea, s/n, E-38200 - La Laguna, Tenerife, Spain\\
              \email{vbejar@iac.es}
	 \and
	     Isaac Newton Group of Telescopes, Apartado de correos 321, E-38700 Santa Cruz de la Palma, Tenerife, Spain\\
	     \email{mischa@ing.iac.es}
	 \and
	     European Southern Observatory, Casilla 19001, Santiago 19, Chile\\
	     \email{nhuelamo@eso.org}
	     \and
	     Department of Physics and Astronomy, UCLA, Los Angeles, CA-90095-1547, U.S.A\\
             \email{ghez@astro.ucla.edu}
	     \and 
	     University of Central Florida, Department of Physics, PO Box 162385, Orlando, FL 32816-2385, USA
}

   \date{Received ; accepted }

   \abstract{We report the initial results of a VLT/NACO high spatial resolution imaging survey for multiple systems among 58 M-type members of the nearby Upper Scorpius OB association. Nine pairs with separations below 1\arcsec\, have been resolved. Their small angular separations and the similarity in the brightness of the components ($\Delta$Mag$_{K}<$1 for all of them), indicate there is a reasonable likelihood several of them are true binaries rather than chance projections. Follow-up imaging observations with WHT/LIRIS of the two widest binaries confirm that their near-infrared colours are consistent with physical very low-mass binaries. For one of these two binaries, WHT/LIRIS spectra of each component were obtained. We find that the two components have similar M6-M7 spectral types and signatures of low-gravity, as expected for a young brown dwarf binary in this association. Our preliminary results indicate a possible population of very low-mass binaries with semimajor axis in the range 100~AU--150~AU, which has not been seen in the Pleiades open cluster. If these candidates are confirmed (one is confirmed by this work), these results would indicate that the binary properties of very low-mass stars and brown dwarfs may depend on the environment where they form.

   \keywords{Very low mass stars, Brown dwarfs, multiple systems
               }
   }
   \authorrunning{Bouy et al.}
   \titlerunning{Multiplicity of M-dwarfs in Upper Scorpius} 
   \maketitle
%

\section{Introduction}
Multiple systems and their physical properties are important testimonies of the formation and early stages of evolution of a class of astrophysical objects. For that reason, the study of multiplicity among ultracool dwarfs has been an intense field of research over the last few years \citep[see e.g][]{1999AJ....118.2460B,1999Sci...283.1718M,2000ApJ...529L..37M,2003ApJ...594..525M,2003AJ....125.3302G,2004A&A...424..213B,2003AJ....126.1526B,2003ApJ...586..512B,2005Natur.433..286C,2003ApJ...587..407C,2002ApJ...567L..53C,2003IAUS..211..233J,2005astro.ph..9134J,2005ApJ...633..452K,2005ApJ...621.1023S}. While all these studies have reported a cut-off in the distribution of separation at about 30~AU, consistent with a dynamical formation mechanism involving gravitational interactions, a still very small but increasing number of wide binary candidates are reported in the field \citep{1998ApJ...509L.113M, 2005A&A...439L..19P,2004A&A...427L...1F,2004AJ....128.1733G,2005A&A...440L..55B} and in young associations \citep{2004ApJ...614..398L,2005A&A...438L..25C,2005A&A...435L..13N}. Together with the relatively high frequency of multiple systems reported by the above mentioned authors, these wide multiple systems challenge the ejection models and show that this scenario, although certainly at work, can probably not explain the formation of the majority of the very low mass objects.

A recent study performed by \citet{2005ApJ...633..452K} in Upper Scorpius has revealed a significant fraction of visual binaries among late type members of the association. The three candidates they resolve among their twelve targets have small separations (less than 18~AU) and flux ratios close to unity, thus similar to the overall field and Pleiades binary populations. More recently \citet{2005ApJ...633L..41L} reported a wide binary (at 130~AU) discovered serendipitously during a spectroscopic survey. The existence of such a fragile wide pair at the age of Upper Scorpius \citep[5~Myr][hereafter Usco]{2002AJ....124..404P}, suggests that some very low-mass objects do not form via ejection. A detailed knowledge of the properties of wide mutliple systems in different environments would provide key constraints regarding the contribution of ejection to the formation of very low mass objects. In this paper, we report the initial results of a systematic search for multiple systems among the very low mass stars and brown dwarfs members of the young Upper Scorpius OB association with the NACO adaptive optics system on the VLT. In section \ref{observations}, we present the observations and processing of the data. In section \ref{dataanalysis} we describe the analysis of the data and the discovery of multiple system candidates. In section \ref{comp} we discuss the companionship of the candidates, and in section \ref{discussion} we discuss the results in the context of the current models of formation and in comparison with previous studies.

\section{Observations and data analysis \label{observations}}

The sample of 58 Upper Scorpius targets was selected in order to cover most of the M spectral class, from M0 to M8, including approximately 5 objects per spectral subtype. This sample can be divided in two parts. The first part, thirty-five targets between M0 and M6, has been selected randomly in the complete sample of \citet{2002AJ....124..404P}. We took care not to select these targets on their luminosity, in order to have a sample statistically representative of the overall population. Almost all \citet{2002AJ....124..404P} objects had K~mag brighter than the K=13~mag limit of the infrared wave front sensor of NACO, making the random selection possible. The second part, consisting of 23 targets in the substellar regime (M5.5 to M8), has been selected among \citet{2004AJ....127..449M} sample. The DENIS 100\% completeness limit is K$\sim$13.5~mag \citep{1997Msngr..87...27E}, but  we selected the targets from this sample according to their magnitude, with a limit corresponding to the limit of sensitivity of the infrared wave front sensor of the NACO adaptive optics ($K<$13~mag). This part of the sample was therefore limited in magnitude. In both cases all targets were confirmed members of the association by the respective discoverers, either via astrometric and/or spectroscopic measurements. In the present paper, we focus on the discovery of the closest (separation less than 1\arcsec) and therefore best candidates, and on the important discovery of a significant number of wide candidates. The detailed statistical analysis of the properties of multiplicity over the whole sample will be presented in a future paper once second epoch data and confirmation of the companionship have been obtained.

\begin{figure}
   \centering
   \includegraphics[width=0.9\textwidth]{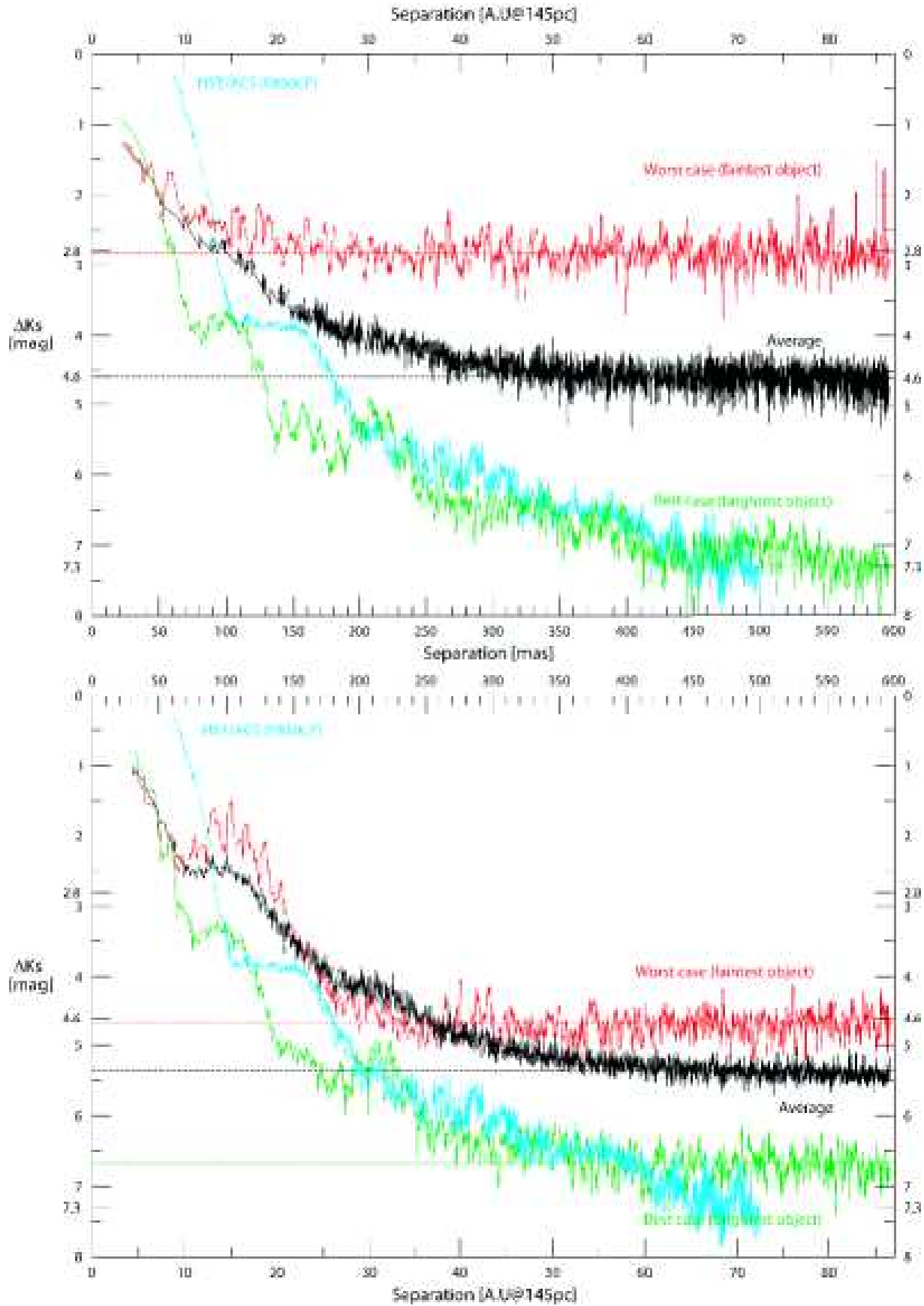}
      \caption{Sensitivity of our NACO observations, computed as the 3-$\sigma$ standard deviation of the radial profile of the PSF. Three cases are represented for the first (top) and second (bottom) nights: the sensitivity curve for the faintest (red) and brightest objects (green), as well as the average of all sensitivity curves (black). The sensitivity of the HST/ACS images of \citet{2005ApJ...633..452K} for Usco brown dwarfs in the F850LP filter is represented as well for comparison (cyan). At close separation, diffraction limit on an 8~m telescope wins. Moreover, our targets having their peak of emission in the near-infrared, a given mass ratio corresponds to a smaller difference of magnitude in Ks than in F850LP, so that the NACO Ks images were sensitive to much lower mass companions than the F850LP ACS images. }
         \label{sensitivity}
   \end{figure}

Table \ref{results} gives a summary of the observations. All targets were observed between 03/06/2005 and 04/06/2005 at the ESO  VLT with its NACO adaptive optics platform \citep{2003SPIE.4841..944L} on UT4. Our targets being faint and red, we used the near-IR wavefront sensor. Standard jitter imaging mode using random pattern with 2 to 20 points (see Table \ref{results}) was used in order to optimize cosmic rays/bad pixels rejection and sky computation. The observational strategy was adapted in real-time to the ambient conditions and the luminosity of the targets in order to optimize the quality of the final images. All objects have been observed in Ks, and one object (DENIS161833) in H band as well. Faint objects were preferably observed at low airmass, and under better seeing conditions, allowing for longer (but fewer) exposures well suited to their faintness, and with more jitter positions. Bright objects were observed at higher airmass and preferably when the ambient seeing was less good, using less jitter positions and shorter (but more) exposures well suited to their higher brightness, and in order to limit the effect of the degraded/variable conditions. 
The individual dithered frames have been checked both automatically using the Eclipse \emph{strehl} task \citep{1997Msngr..87} and by eyes in order to select only the best images, and optimize the quality of the final image. The selected frames were then average combined using the recommended Eclipse \emph{jitter} task.  
The atmospheric conditions during the two nights were ideal, allowing us to obtain sharp images, even for the faintest of our targets. The strehl ratios of the final images range between 13\% and 35\%. Figure \ref{sensitivity} shows the limit of sensivity of our survey during the two nights. It ranges from $\Delta$Mag=1~mag at 0\farcs030 (almost independently from the brightness of the targets) to $\Delta$Mag=2.8, 5 and 7.3~mag at 0\farcs4 and greater, for the faintest, average luminosity and brightest targets respectively.

Complementary near-infrared images of two binary candidates could be obtained with LIRIS \citep{1998SPIE.3354..448M} at the William Herschel Telescope in La Palma. LIRIS uses a 1024$\times$1024 HAWAII near-infrared detector, with a pixel scale of 0\farcs25/pixel, yielding a field of view of 4\farcm27 x 4\farcm27. The two observations (USCO-160028.4 and DENIS161833) were performed respectively on 2005 June 18th and 19th, the first one in both J and H, and the second one in J only. Standard dithering mode using a 5-point pattern with individual exposure times of 10~sec was used in order to optimize cosmic rays/bad pixels rejection and sky computation. The corresponding images were then processed using the \emph{THELI} image reduction package \citep{2005AN....326..432E}. In addition, we also obtained a near-IR spectrum of DENIS161833, with exposures times of 6$\times$200~sec and 6$\times$120~sec in $ZJ$ and $HK$ respectively. The 2-D spectra have been processed using standard procedures in \emph{IRAF}\footnote{IRAF is distributed by the National Optical Astronomy Observatory, which is operated by the Association of Universities for Research in Astronomy, Inc., under cooperative agreement with the National Science Foundation.}. The individual spectra were then extracted using the same method as described in \citet{2004A&A...423..341B}, fitting the line spread function with a gaussian. The correction for the atmospheric extinction was applied using the spectrum of a B9V standard star observed immediately after the target.  The various hydrogen lines of the spectrum of the hot star have been removed manually. The spectra are presented in Figure \ref{spectra}. The final resolution of the NIR spectra is as follows: 20~\AA\,  in $ZJ$, and 33~\AA\, in $HK$. The choice of a slit much narrower than the seeing does not allow us to perform flux calibration using the standard star, but we could correct the relative slope between $ZJ$ and $HK$ using the broad-band J, H and K photometry given in table \ref{results}.

\section{Analysis of the images \label{dataanalysis}}

\begin{figure}
   \centering
   \includegraphics[width=0.8\textwidth]{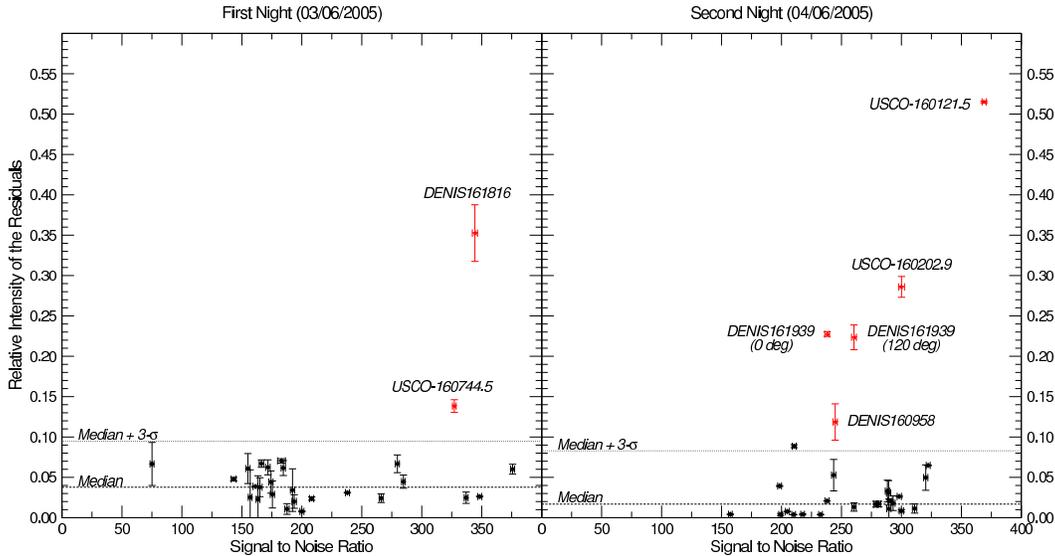}
      \caption{Relative intensity of the residuals after PSF subtraction. An object is considered as binary candidate when the residuals are higher than 3-$\sigma$ above the median value of the sample. The nine candidates presented in this paper are clearly detected (indicated in red). One object is classified as unresolved although it falls just above the 3-$\sigma$ limit on the second night, because a detailed visual inspection reveals that it is almost certainly due to the triangular aberations (variable and typical for NACO PSFs) stronger than usual in that particular image.}
         \label{residuals}
   \end{figure}

\begin{figure}
   \centering
   \includegraphics[width=0.6\textwidth]{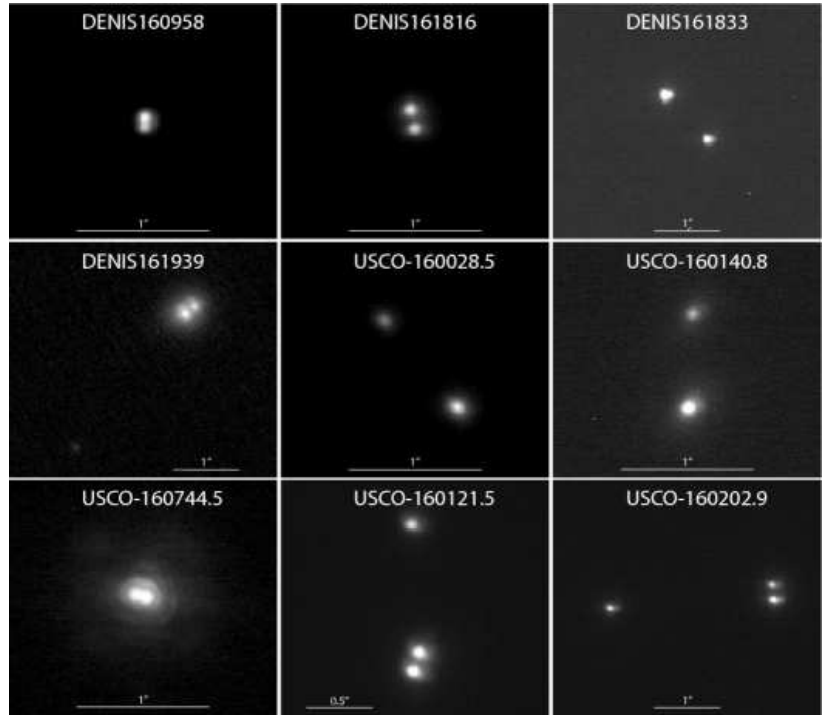}
      \caption{Mosaic of NACO images of all candidate multiple systems with separation less than 1\farcs0. The scale is indicated. North is up and east is left.}
         \label{mosaic}
   \end{figure}

We identify close\footnote{separation less than $\sim$1\arcsec} companion candidates either by eye in the case of well resolved systems, or using the method described in \citet{2005AJ....129..511B}. Briefly, it consists in a systematic analysis of the residuals after PSF subtraction. Three unresolved reference PSFs obtained just before and three other obtained just after each target were used to perform the PSF subtraction and the analysis of the residuals. Moreover, we made sure to select reference PSFs with similar spectral types and magnitudes, in order to have similar AO corrections and PSF properties. Nine companions with separations less than 1\arcsec\, were unambiguously identified with these methods, as shown in Figure \ref{residuals}. Three of the multiple system candidates also have a third object in the field of view (see Figure \ref{mosaic}), at less than 2\farcs6, making them relatively good triple system candidates (see section \ref{comp} for a detailed discussion on the multiplicity). 

The relative astrometry and photometry of the close candidate multiple systems have been obtained on the NACO images using the dual-PSF fitting program described in \citet{2003AJ....126.1526B}, and adapted to NACO. Briefly, the program performs a dual-PSF fit of the binary, fitting both component at the same time. The relative astrometry and photometry are obtained when the residuals reach their minimum value. The program repeats the fit using 6 different reference PSFs chosen as described previously. The final value and uncertainties are obtained by averaging the 6 results. Table \ref{results} gives an overview of the results. The relative astrometry and photometry of the wider companion candidates (separation greater than 1\arcsec) were measured using the IRAF \emph{phot} task, using an aperture radius of 35 pixels corresponding to $\sim$5 times the FWHM. The absolute photometry of the individual components was then calculated using the 2MASS magnitude of the unresolved pairs and the flux ratios measured with the PSF fitting. USCO-160028.4 and DENIS161833 are just resolved in the LIRIS images. The same PSF fitting program as described previously is used to measure the relative astrometry and photometry in the J and H bands, with reference PSFs taken in the field of view of the image. 

\section{Companionship \label{comp}}

\subsection{Contamination by foreground/background coincidences}
\begin{figure}
   \centering
   \includegraphics[width=0.6\textwidth]{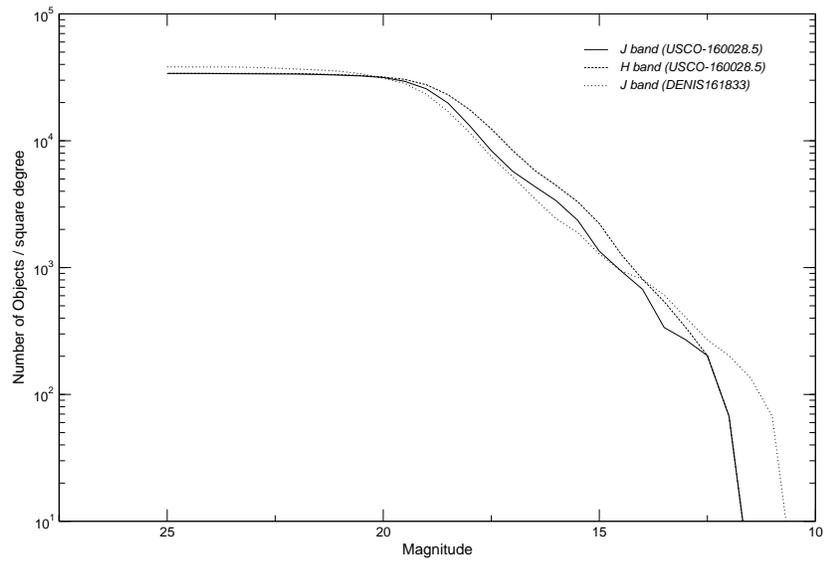}
      \caption{Cumulative brightness function $\rho(m)$ from the LIRIS 4\farcm7$\times$4\farcm7 J and H-band images, scaled to 1~deg$^{2}$.}
         \label{surf_dens}
   \end{figure}

It is hard to assess the companionship of the candidates from the NACO Ks band images only. Spectroscopic observations of the individual components and second epoch measurements are required to conclude on the binarity of the candidates presented here.
One way to estimate the contamination by unrelated visual companions is to estimate the surface density of objects. We used the three wide-field J and H-band LIRIS images. Although obtained in a different filter than the NACO images, their sensitivities are higher by about 2~mag than the most sensitive of our NACO Ks image, ensuring a conservative  measurement of the surface density. Figure \ref{surf_dens} shows the cumulative brightness function $\rho(m)$ of unrelated background sources measured in the 3 images and scaled to 1 deg$^{2}$. This figure shows a good agreement with the results obtained by \citet{2000AJ....120..950B} with HST/NICMOS in the F110W filter for sources in Scorpius Centaurus. As described in this latter paper, the probability $P(\Theta,m)$ for an unrelated source to be located within an angular distance $\Theta$ from a particular target is given by:
$$P(\Theta,m)=1-e^{-\pi\rho(m)\Theta^{2}}$$
According to this equation, the probability for our faintest companion candidate (USCO-160028.5B, J=14.5~mag) being an unrelated background source at 0\farcs9 is only $\sim$0.1\%. A similar reasoning would give very low probabilities for the other candidates as well. Such a calculation is only tentative and must be considered with caution. First because it uses only two particular fields of Upper Scorpius, while it is well known that the density of objects (members and non-members) can vary greatly within the association. Second because several studies have shown in the past that objects considered as physical pairs on the basis of similar calculations finally turned out to be unrelated objects after follow-up observations. Spectroscopy of the individual components, as well as common proper motion, are required in order to confirm the multiplicity of the candidates.

The complementary LIRIS observations provide useful informations to assess the multiplicity of USCO-160028.4 and DENIS161833

\subsection{DENIS161833}
\begin{figure}
   \centering
   \includegraphics[width=0.8\textwidth]{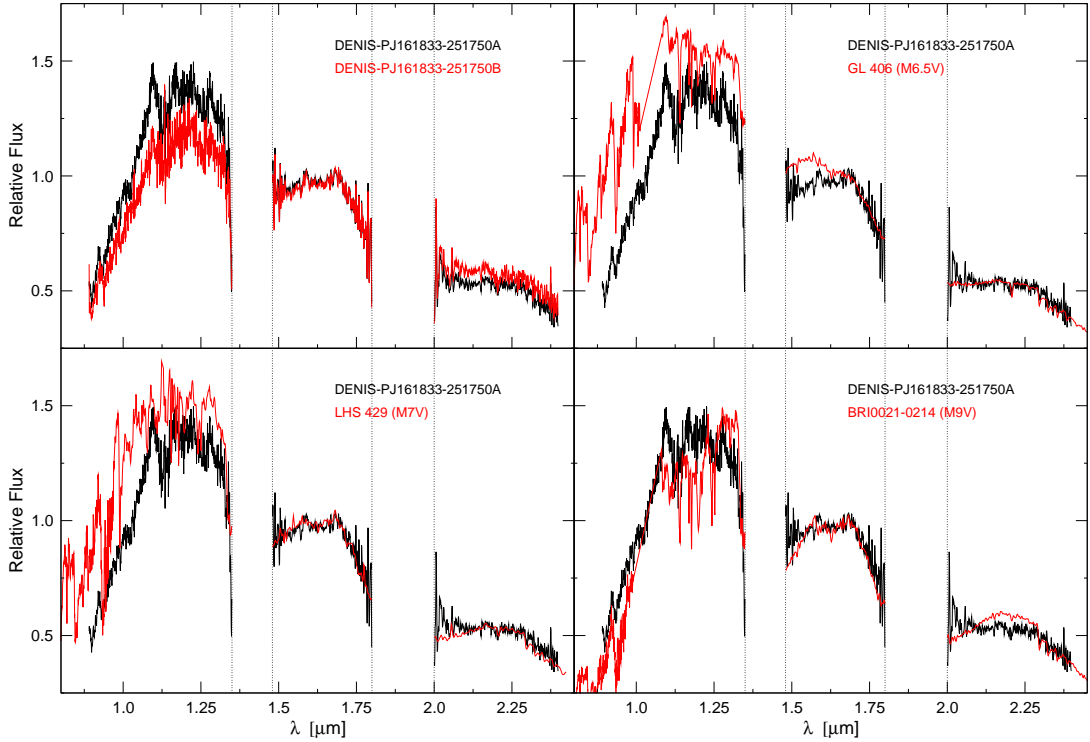}
      \caption{LIRIS $ZJ$ and $HK$ spectra of the 2 components of DENIS161833. The two objects clearly have similar spectral types (top-left panel). The spectrum of the primary is compared to spectra GL406 (M6.5, top-right panel), LHS~429 (M7, bottom-left panel) and BRI0021-0214 (M9, bottom-right panel). Theb est match in the H-band is obtained with the M7 dwarf.}
         \label{spectra}
   \end{figure}

\begin{figure}
   \centering
   \includegraphics[width=0.3\textwidth]{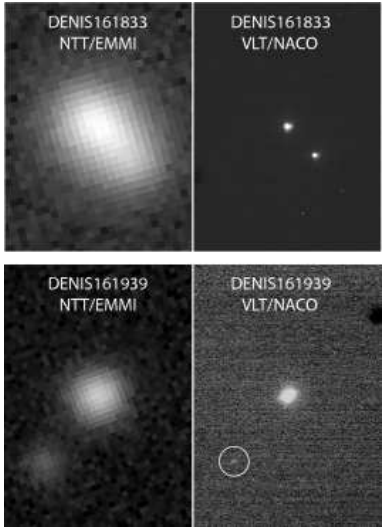}
      \caption{Comparison of NTT/EMMI I-band (left) and VLT/NACO Ks-band (right) images of DENIS161833 (top) and DENIS161939 (bottom). Orientation and scales between EMMI  and NACO  images are matched for each object. DENIS161833 is clearly elongated in the EMMI image. The position angle and separation look consistent, but it is not possible to measure an accurate relative position of the two components. DENIS161939C is clearly resolved in the EMMI image. It is indicated with a circle in the NACO image. The separation and position are consistent with the NACO images, but the poor accuracy does not permit to conclude that there is common proper motion. The third component appears to be much brighter in I (EMMI) than in K (NACO). Such a blue object is very likely to be unrelated to the primary.}
         \label{emmi}
   \end{figure}

\citet{2004AJ....127..449M} classified DENIS161833 as M6 and as member in the association on the basis of their location in the magnitude vs. spectral type and $W_{Na I}$ vs spectral type diagrams. Table \ref{results} shows that the two components of DENIS161833 have near-infrared magnitudes and colours consistent with two objects of similar spectral class. Figure \ref{spectra} shows the LIRIS spectra of DENIS161833A and B. We compare the two objects one with each other but also with spectra of M6.5V, M7V and M9V dwarfs from \citet{2001ApJ...548..908L,2000ApJ...535..965L}. The photometric uncertainties in the $ZJ$ band being significantly larger, we use the H-band spectra to perform the comparisons.  Figure \ref{spectra}  shows that the two spectra have almost identical shapes, indicating that the two components must belong to very close spectral subtypes. The B component must be half a subclass to one subclass cooler than A based on the slightly larger water vapor absorption in the H-band and the larger flux in the K band.  The K~I absorption lines appear to be weaker in both DENIS161833A and B than in the field objects, indicating lower gravity atmospheres. This is consistent with  the low gravity already reported by \citet{2004AJ....127..449M} from the unresolved optical spectrum, and confirms the likely membership of the two components to the association. The spectrum of DENIS161833A best matches that of an M7 dwarf, slightly cooler than the spectral type of M6 reported by \citet{2004AJ....127..449M} for the unresolved system. We also retrieved the pre-acquisition NTT/EMMI frame obtained by the support astronomer in service mode before the optical spectrum of DENIS161833 presented in \citet{2004AJ....127..449M}. These images were obtained in the I band with exposure time of only 10~s. No calibrations were obtained so that we could not process the image. The seeing was not good enough ($\sim$1\arcsec) and the target is not resolved but clearly elongated (see Figure \ref{emmi}). We tried to measure the precise position angle and separation using the same PSF fitting method as described above, but because of  the poor quality of the unprocessed data, the program never converged to acceptable values. Although the position angle and separation look consistent at a first look, an accurate analysis of common proper motion is not possible with these images and no conclusion can be drawn from these images.

From their 6~deg$^{2}$ survey, \citet{2002AJ....124..404P} identified 62 M-dwarfs members, leading to a density of M0--M6 dwarfs of 1.1~deg$^{-2}$. Although this value is only an upper limit, it shows that the probability to find a target belonging to the association and with a similar spectral type within 1\arcsec\, must be extremely low. We therefore consider that DENIS161833AB is confirmed as an Upper Scorpius brown dwarf binary at a very high level of confidence. Second epoch images in the coming years should confirm common proper motion.

At the time this paper was submitted, \citet{2005ApJ...633L..41L} presented independent imaging and spectroscopic observations obtained 2 weeks after ours leading to the same conclusions as the ones we draw here, and assessing the multiplicity of this object.

\subsection{USCO-160028.4}
\citet{2002AJ....124..404P} classified the unresolved pair as M6 and as member of the association based on the presence of a relatively strong lithium absorption line in its spectrum. They also estimate an absorption A$_{V}$ of 0.0. The two components of USCO-160028.4 have similar near-infrared magnitudes and colours (both have J-K$\sim$0.8 and H-K$\sim$0.3, see Table \ref{results}), consistent with objects of similar spectral type, and increasing the probability that the two objects are physically bound. The probability to find two objects with similar luminosities and near-IR colours within less than 1\arcsec\, is indeed extremely low, as discussed previously. Common proper motion and spectroscopy of the two components are required to confirm the multiplicity.

\subsection{DENIS161939C \label{denis161939}}
We also retrieved the pre-acquisition NTT/EMMI frame obtained before the optical spectra of DENIS161939 presented in\citet{2004AJ....127..449M}. As in the case of DENIS161833, the lack of calibration frames did not allow us to process the image. The seeing was good enough ($\sim$1\arcsec) to resolve the third component of the candidate triple system (see Figure \ref{emmi}). Again, the PSF fitting method did not converge to any satisfactory results. The objects beeing well resolved, we could use standard centroiding to measure the separation (2\farcs662$\pm$0\farcs023) and position angle (141.8\degr) of the third component. The two measurements are consistent but the 1-$\sigma$ uncertainties being of the same order than the mean proper motion of the association \citep[$\mu_{l} \cos{b}$=-24.5$\pm$0.1~mas/yr, and $\mu_{b}$=-8.1$\pm$0.1~mas/yr, ][]{1999AJ....117..354D} we cannot draw any firm conclusion regarding common proper motion. We tentatively measure the relative photometry using an aperture radius of 8 pixels (1\farcs328). We measure a difference of magnitude in the I band of $\Delta$Mag(I)=2.05~mag. Uncertainties on this value must be large since we used the raw images, but the third component appears to be certainly much bluer than the primary, which indicates that the third component is likely to be a background/foreground coincidence.

\subsection{Comparison with the HST study \label{hst}} 

Using HST/ACS, \citet{2005ApJ...633..452K} resolved 3 companion candidates in a sample of 12 brown dwarfs in USco. All 3 of them have separations closer than 18~AU. The observed frequency of tight binaries found in the HST study (3/12=25$^{+15}_{-8}$\%) is consistent within the error bars with our observed binary frequency in the same separation range (0--2175~AU in average for our NACO study) and spectral type range (M5.5--M7.5) . The wide companion candidates that they observed in their ACS frames were rejected as background coincidences on the basis of their position in an optical colour-magnitude diagram ($i'$ vs $i'-z'$). How does this non detection of wide binaries by HST compare with our current results in a larger sample?

Although the filters and sensitivities of the two surveys are different, the candidates having flux ratios close to unity make the comparison meaningful, since they would have been detected in both surveys. Among our 28\footnote{23 targets from \citet{2004AJ....127..449M} and 5 from \citet{2002AJ....124..404P}} targets with spectral types within the range covered by the \citet{2005ApJ...633..452K} survey (M5.5--M7.5), we resolve two wide binary candidates (one is confirmed and the other has consistent near-IR colours), leading to an observed fraction of 7$_{-2}^{+8}$\%. This wide binary frequency implies that \citet{2005ApJ...633..452K} should have detected 0.85 wide binaries, which is consistent with their non detection. We can combine the HST and NACO studies to give an observed wide binary frequency in USco of 2/40 (5$_{-2}^{+6}$\%) for primary spectral types in the range M5.5--M7.5. 

\section{Discussion \label{discussion}}

Second epoch imaging and spectroscopic observations of the candidates are required before drawing any firm conclusion, but we can tentatively discuss the properties of multiplicity of our sample of candidates in comparison with field and Pleiades objects. The following discussion is therefore only preliminary, until new observations confirm the companionship of the candidates (and in particular of the wide candidates). 

\begin{figure}
   \centering
   \includegraphics[width=0.6\textwidth]{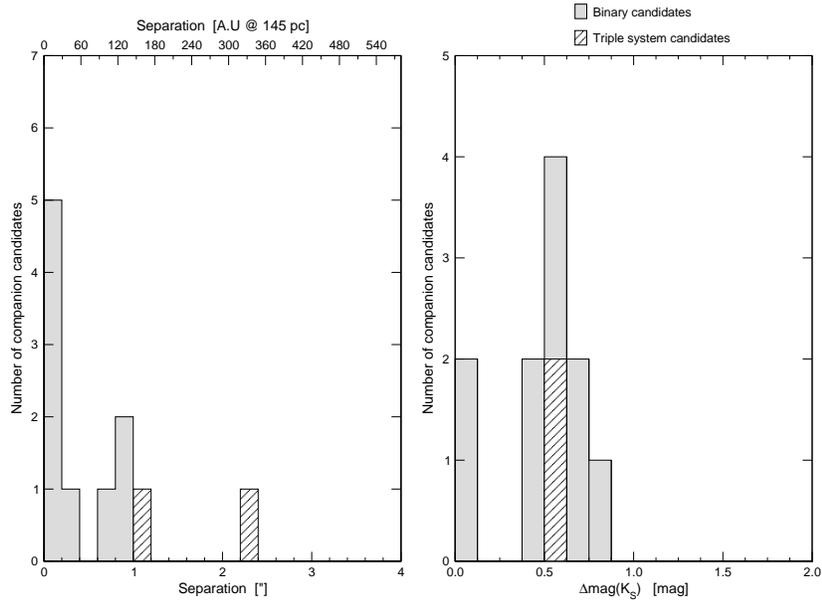}
      \caption{Distribution of separation (left panel) and of difference of magnitude (right panel) of the companion candidates. The grey part of the histogram represents the measurements for the binary candidates, and the hashed part represents the measurements for the third component of the triple system candidates. The NACO images were limited to $\sim$0\farcs030 at close separations, and $\sim$13\farcs0 at large separations, and to $\Delta$mag$\sim$2.8~mag at worst, $\sim$7.3~mag at best and $\sim$5 in average.}
         \label{distrib}
   \end{figure}

Figure \ref{distrib}  shows that two third (6/9) of the candidates have a separation less than 30~AU, consistent with the field or Pleiades objects observations, while the last third  (3/9) have separations greater than 100~AU, contrasting with field or Pleiades objects where only few wide multiple systems have been reported to date \citep{1998ApJ...509L.113M, 2005A&A...439L..19P,2004A&A...427L...1F,2004AJ....128.1733G,2005A&A...440L..55B}. However, when combining our results with the HST study of \citet{2005ApJ...633..452K} (see section \ref{hst}), we obtain an observed wide binary frequency in USco of 2/40 (5$_{-2}^{+6}$\%) for primary spectral types in the range M5.5--M7.5, which could be consistent with the tail of the separation distribution reported for field objects in that spectral type range citep[see e.g][]{2003ApJ...587..407C,2003AJ....126.1526B,2003AJ....125.3302G}. Two binary candidates (over nine) have a third component candidate. These third components are the widest companion candidates, which could be explained by the higher binding energy from their binary companion. This fraction of higher order multiple systems (2/9) and their separations contrasts with field objects, where only two close triple candidates have been reported to date \citep{2005AJ....129..511B,martin_stis}.

The distribution of separation of the candidate multiple systems we report is similar to the properties of the binary candidates reported by \citet{2000AJ....120..950B} from their HST/NICMOS survey among a sample of young, X-Ray active late-type stars in the Scorpius-Centaurus association. The separations of their six M0--M3 Scorpius-Centaurus multiple system candidates indeed range between 0\farcs05 and 2\farcs9, but the companion candidates have not been confirmed to date and they evaluate a high probability of contamination by background sources. 


If confirmed by second epoch observations, these results will challenge the models of formation involving ejection at the early stages, since ejection alone cannot produce such a significant number of wide binaries (3/9) and of triple systems (2/9). Numerical simulations by \citet{2002MNRAS.332L..65B} have indeed recently shown that the formation of brown dwarfs and very low mass stars via ejection from the protostellar cluster lead to a clear preference for close binaries, even though some rare wide ejected systems can be formed by pairing after nearly simultaneous ejections. Finally, these models should also reproduce the timescale (within less than 100~Myr) and the cut-off (at about 20--30~AU) observed after disruption of the unstable wide systems, as observed in the Pleiades and in the field. Considering the dynamical evolution of 3-body systems, \citet{2005ApJ...623..940U} were able, under specific initial conditions and assumptions, to reproduce the observed distribution of separation observed for old field objects. If, as it is thought, most objects are born in OB associations like Upper Scorpius \citep[e.g][]{1978PASP...90..506M}, the apparent discrepancy between old isolated field M-dwarfs and their young Upper Scorpius counterparts would indicate that dynamical evolution must be still at work at the age of Upper Scorpius ($\sim$5~Myr). While they were sensitive to this separation range, none of the successive high angular resolution surveys performed in the Pleiades with HST \citep{2000ApJ...543..299M,2003ApJ...594..525M,2005pleiadesinprep} have found companion candidates in the range 100-130~AU. The lack of wide binaries among Pleiades very low mass stars and brown dwarfs is therefore real, but it is hard to know whether the difference with Upper Scorpius binary properties is due to the denser environment, where gravitational encounters must have been more frequent, or as in the field to the disruption of the weakly bounded wide systems. From their survey in Lupus~3, \citet{2005A&A...440..139L} report two brown dwarfs candidates located at 560~AU and 840~AU of previously known Lupus stars,  but these candidates have been identified only on the basis of their projected proximity in the sky and are very likely to be unrelated visual pairs. 

Figure \ref{distrib}  shows also that all binary candidates have differences of magnitudes in Ks less than 0.9~mag, corresponding to a mass ratio of about 40\% according to the DUSTY evolutionnary models of \citet{2002A&A...382..563B} at an age of 5~Myr. This preference for equal mass systems is real, since we were sensitive to differences of magnitude of 2.8~mag in the worst case, and 7.3~mag in the best case, corresponding to mass ratios of about 15\% and 3\% respectively. It is also reported by \citet{2005ApJ...633..452K} for their 3 binary candidates, and from \citet{2000AJ....120..950B} for their 6 M0--M3 candidates (with differences of magnitude between 0.5--1.7~mag in the F108N NICMOS filter). This is similar to the tendency observed up-to-now in the field and in the Pleiades, although a direct comparison is strictly not possible since the field and Pleiades samples were strongly biased towards equal mass systems. Surveys in even younger associations should allow to understand whether these systems form and are quickly destroyed, or do not form at all. 

Among the three triple systems, two have differences of magnitude of about 0.5~mag, making them similar to the binaries, and therefore more likely to be physically bound. The multiplicity of the last one (DENIS161939) has already been discussed in see section \ref{denis161939}. 

\section{Conclusions}
From our high angular resolution survey among a sample of 58 ultracool dwarfs in Upper Scorpius, we report 9 new binary candidates, among which 2 might be triples. Complementary near-infrared imaging and spectroscopic observations indicate that one of the wide multiple systems (DENIS161833AB) is a physical pair of M7--M7.5 brown dwarfs at a high level of confidence. Independent observations by \citet{2005ApJ...633L..41L} lead to the same conclusion. Second epoch imaging and spectroscopic observations are required to confirm the multiplicity of the eight other candidates, but the near-infrared colours of one of them (USCO-160028.5AB) are consistent with a companion of similar spectral type. All binary candidates have differences of magnitude corresponding to mass ratio close to unity, confirming a preference for equal mass systems similar to what was observed up-to-now for field and Pleiades objects. The distribution of separations is extending up to separations of 100~AU, contrasting with the cut-off observed at 30~AU for field and Pleiades ultracool objects. With one third (3/9) of the candidates having separations greater than 100~AU, two third (6/9) in the range of separation between 0--30~AU, and two triple systems, the properties of the new candidates contrast with the properties of multiple ultracool dwarfs from the field or from the Pleiades, but the multiplicity of 8 of them must be confirmed before any firm conclusion can be drawn.

\begin{acknowledgements}
We are very grateful to our nighttime support astronomer in Paranal Olivier Marco, whose excellent work made it possible to obtain the data presented in this paper, to our daytime support astronomers, Dominique Naef and Christophe Dumas, as well as to the Paranal SciOps team in general for their kind and efficient support. H. Bouy thank Jos\'e Acosta Pulido for his advices and help in the reduction of the LIRIS spectra.

\end{acknowledgements}

\bibliographystyle{aa}

\begin{deluxetable}{lllllllllll}
\tablecaption{Observation log and relative astrometry and photometry of the candidate multiple systems \label{results}}
\tabletypesize{\scriptsize}
\tablewidth{0pt}
\tablehead{
\colhead{Name} & \colhead{R.A (2000)} & \colhead{Dec. (2000)} & \colhead{SpT} & \colhead{Exp. Time} & \colhead{Sep.} & \colhead{P.A [\degr]} & \colhead{$\Delta$mag} &  Mag(A) &\colhead{Comments} 
}
\startdata
DENIS161833AB	& 16 18 33.2 & -25 17 50.4 &  M7   & 2x45  & 0\farcs924$\pm$0\farcs002 & 223.6  &  Ks=0.62$\pm$0.08 & 11.71 &     \\
		&	     &  	   &	   & 2x45  &	                       &        &  H= 0.65$\pm$0.08 & 12.18 & \\
		&	     &  	   &	   & 5x10  & 0\farcs914$\pm$0\farcs025 & 222.7  &  J= 0.80$\pm$0.1  & 12.90 & LIRIS obs.     \\
DENIS160958AB	& 16 09 58.5 & -23 45 18.6 &  M6.5 & 6x45  & 0\farcs080$\pm$0\farcs003 & 173.3  &  Ks=0.41$\pm$0.08 & 12.06 &     \\
DENIS161816AB	& 16 18 16.2 & -26 19 08.1 &  M5.5 & 8x45  & 0\farcs147$\pm$0\farcs003 & 192.3  &  Ks=0.12$\pm$0.03 & 11.59 & \\
USCO-160028.5AB	& 16 00 28.5 & -22 09 22.9 &  M6   & 6x45  & 0\farcs855$\pm$0\farcs003 & 40.0   &  Ks=0.73$\pm$0.08 & 12.97 &  \\
		&	     &  	   &	   & 5x10  & 0\farcs845$\pm$0\farcs025 & 40.5   &  H= 0.74$\pm$0.10 & 13.30 & LIRIS obs.     \\
		&	     &  	   &	   & 5x10  &                           &        &  J= 0.66$\pm$0.10 & 13.84 & LIRIS obs.     \\
USCO-160140.8AB	& 16 01 40.8 & -22 58 10.4 &  M3   & 20x5  & 0\farcs706$\pm$0\farcs001 & 357.5  &  Ks=0.84$\pm$0.03 & 10.26 &  \\
USCO-160744.5AB	& 16 07 44.5 & -20 36 03.1 &  M4   & 18x10 & 0\farcs077$\pm$0\farcs003 & 73.1   &  Ks=0.64$\pm$0.10 & 9.56  &   \\
\hline
\multicolumn{10}{c}{Candidate triple systems} \\
\hline
DENIS161939AB	& 16 19 39.8 & -21 45 35.1 &  M7   & 3x75  & 0\farcs177$\pm$0\farcs003 & 309.7  &  Ks=0.55$\pm$0.08 & 12.61 &  \\
DENIS161939AC	&            &             &       &       & 2\farcs598$\pm$0\farcs003 & 140.8  &  Ks=3.88$\pm$0.28 & 12.61 & unrelated \\
USCO-160121.5AB & 16 01 21.5 & -22 37 26.5 &  M4   & 11x30 & 0\farcs146$\pm$0\farcs001 & 343.5  &  Ks=0.04$\pm$0.03 & 11.08 & \\
USCO-160121.5AC &            &             &       &       & 1\farcs116$\pm$0\farcs001 & 0.8    &  Ks=0.54$\pm$0.02 & 11.08 & \\
USCO-160202.9AB	& 16 02 02.9 & -22 36 14.0 &  M0   & 5x75  & 0\farcs227$\pm$0\farcs001 & 5.9    &  Ks=0.47$\pm$0.03 & 12.50 &  \\
USCO-160202.9AC	&            &             &       &       & 2\farcs458$\pm$0\farcs001 & 92.9   &  Ks=0.52$\pm$0.05 & 12.50 &  \\
\enddata
\tablecomments{Unless specified, the value refer to NACO observations. Spectral Types of the unresolved systems from \citet{2002AJ....124..404P} and \citet{2004AJ....127..449M}}. 
\end{deluxetable}

\end{document}